\documentclass[superscriptaddress]{revtex4-2}
\usepackage{graphicx}

\usepackage{pdfpages}
\usepackage{etoolbox} 

\makeatletter
\patchcmd{\@outputpage@head}{\@ifx{\LS@rot\@undefined}{}{\LS@rot}}{}{}{}

\begin{document}

\title{Ultrafast dynamics in the Lifshitz-type 5${d}$ pyrochlore antiferromagnet Cd$_{2}$Os$_{2}$O$_{7}$}

\author{Inho Kwak}
\author{Min-Cheol Lee}
		\altaffiliation{Current affiliation: Center for Integrated Nanotechnologies, Los Alamos National Laboratory, Los Alamos, New Mexico 87545, USA}
\author{Byung Cheol Park}
		\altaffiliation{Current affiliation: Department of Energy Science, Sungkyunkwan University, Suwon 16419, Repubolic of Korea}
\author{Choong H. Kim}
\author{Bumjoo Lee}
	\affiliation{Center for Correlated Electron Systems (CCES), Institute for Basic Science (IBS), Seoul 08826, Republic of Korea}
	\affiliation{Department of Physics and Astronomy, Seoul National University, Seoul 08826, Republic of Korea}
\author{C. W. Seo}
	\affiliation{Department of Physics, Chungbuk National University, Cheongju, Chungbuk 28644, Republic of Korea}
\author{J. Yamaura}
	\affiliation{Materials Research Center for Element Strategy, Tokyo Institute of Technology, Kanagawa 226-8503, Japan}
\author{Z. Hiroi}
	\affiliation{Institute for Solid State Physics (ISSP), University of Tokyo, Kashiwa 277-8581, Japan}
\author{Tae Won Noh}
		\thanks{Corresponding author}
		\email{twnoh@snu.ac.kr}
	\affiliation{Center for Correlated Electron Systems (CCES), Institute for Basic Science (IBS), Seoul 08826, Republic of Korea}
	\affiliation{Department of Physics and Astronomy, Seoul National University, Seoul 08826, Republic of Korea}
\author{K. W. Kim}
		\thanks{Corresponding author}
		\email{kyungwan.kim@gmail.com}
	\affiliation{Department of Physics, Chungbuk National University, Cheongju, Chungbuk 28644, Republic of Korea}
\date{\today}

\begin{abstract}
We investigate the ultrafast dynamics of Cd$_2$Os$_2$O$_7$, a prototype material showing a Lifshitz-type transition as a function of temperature. In the paramagnetic metallic state, the photo-reflectivity shows a sub-picosecond relaxation, followed by a featureless small offset. In the antiferromagnetic state slightly below $T_N$, however, the photo-reflectivity resurges over hundreds of picoseconds, which goes beyond the usual realm of the effective-temperature model. Our observations are consistent with the Lifshitz phase transition of Cd$_2$Os$_2$O$_7$ driven by the evolution of the local magnetic moment.
\end{abstract}

\maketitle

%%%%%%%%%%%%%%%%%%%%%%%%%%%%%%%%%%%%%%%%%%%%%%%%%%%%%%%%%%%%%%%%%%%%%%%%%%%%%%%%%%%%%%%%%%%%%%%%%%%

\textit{I. Lifshitz} showed that a Fermi surface topology of a metal can change during continuous deformation under pressure \cite{Lifshitz1960}. The Lifshitz transition due to various instabilities can be accompanied by a metal-insulator transition (MIT) as a function of pressure or temperature without doping \cite{Lifshitz1960,Shinaoka2012,Volovik2017,Braganca2018}. In addition, various physical properties can exhibit anomalies across the Lifshitz transition \cite{Lifshitz1960,Volovik2017,Braganca2018,Wu2015,Marsik2013,Ren2017,Shi2017,Chatterjee2017,Zhang2017}. In most of cases, however, Lifshitz transitions have been obscured by the presence of other bands at the Fermi level across the transition and/or extrinsic effects due to doping \cite{Wu2015,Marsik2013,Ren2017,Shi2017,Chatterjee2017,Zhang2017}. Therefore, various aspects of the Lifshitz transition remain largely unexplored.

The 5$d$ pyrochlore Cd$_2$Os$_2$O$_7$ is a prototype all-in-all-out (AIAO) antiferromagnetic material exhibiting the Lifshitz transition accompanied by MIT. Because its resistivity increases abruptly below the antiferromagnetic transition temperature $T_N$ = 227 K, the MIT is believed to be closely related to the magnetic order. Earlier studies on 5$d$ pyrochlore Cd$_2$Os$_2$O$_7$ claimed a Slater-type MIT that occurs concurrently with unit cell doubling due to the magnetic order \cite{Padilla2002,Mandrus2001}. However, the strong spin-orbit coupling of the osmium 5$d$ orbitals drives AIAO antiferromagnetic ordering \cite{Bogdanov2013,Calder2016,Nguyen2017}. The lack of unit cell doubling in the AIAO spin structure rules out the possibility of a Slater-type transition \cite{Tardif2015,Yamaura2012}. Recent studies have suggested a Lifshitz-type MIT driven by the magnetic ordering. Temperature dependent evolution of the magnetic moment is suggested to push electron and hole bands away from the Fermi level \cite{Shinaoka2012}. Therefore, the system provides a good testbed to explore the temperature dependent evolution of various properties across the Lifshitz transition.

\begin{figure}[]
	\includegraphics[width = 0.45\textwidth]{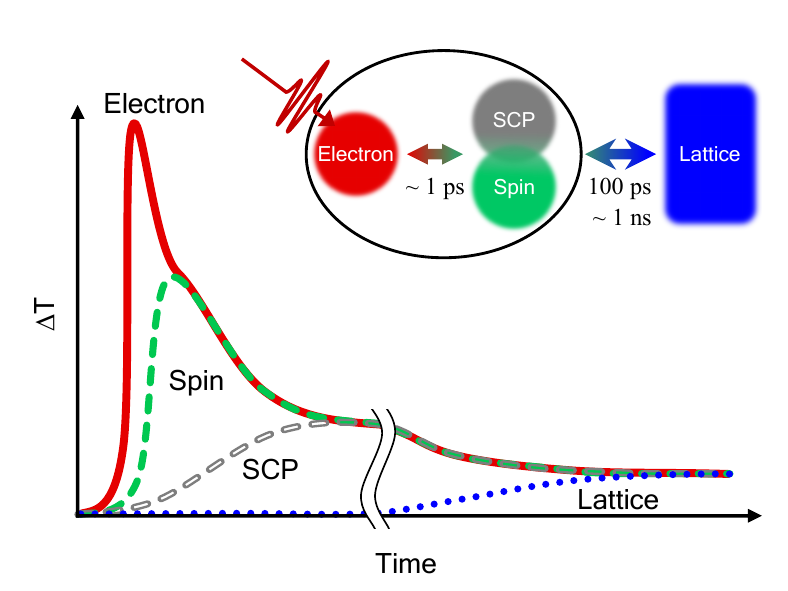}
	\caption{\label{fig:sch}(Color online) Schematic diagrams of the relaxation dynamics in terms of the effective-temperatures of the subsystems. Colored lines indicate the effective-temperature changes of the electron (solid line), spin (filled dashed line), strongly coupled phonon (SCP, open dashed line), and total lattice (dotted line) subsystems. 
	}
\end{figure}

The interactions between quantum degrees of freedom such as charge, spin and lattice have been investigated extensively using ultrafast techniques \cite{Beck2013,Kabanov2005,Allen1987,Groeneveld1995,Perfetti2007,Giannetti2016,Averitt2002,Watanabe2009,Lisowski2005,Patz2014,Gerber2017,Smallwood2012,Kim2012,Gerber2015}. Effective-temperature model provides the simplest picture of non-equilibrium dynamics. Figure \ref{fig:sch} shows a schematic diagram of the relaxation process based on the effective-temperatures of the electron ($T_E$), strongly coupled phonon ($T_{\text{SCP}}$), spin ($T_S$), and total lattice ($T_L$) subsystems. The ultrafast pump excitation generates hot electrons in the excited states, thus elevating $T_E$. Then, the hot electrons cool down by dissipating their excess energy to the other subsystems. The most effective cooling channel is the scattering by the strongly coupled phonons (SCPs), which dominates the sub-picosecond (ps) dynamics \cite{Allen1987,Groeneveld1995,Perfetti2007,Giannetti2016,Averitt2002}. This scattering process can be modified by the magnetic order, which introduces the spin subsystem into the ps region \cite{Averitt2002,Lisowski2005,Watanabe2009,Patz2014}. Investigating such ps dynamics is crucial to understanding the interplay between the electron, SCP, and spin subsystems. We note that $T_L$ remains close to the equilibrium temperature during this ps relaxation. The further heat-transfer from the electron-SCP-spin subsystems to the lattice, i.e., the entire phonon system, takes place via the phonon-phonon or spin-phonon scattering over hundreds of ps or longer. After those scatterings, all the subsystems finally result in a quasi-equilibrium state at a slightly elevated temperature. The total system heating after the equilibration is much smaller than the initial heating of the electron subsystem, because the specific heat of the electron subsystem is only a small proportion of the total specific heat. Hence, the non-equilibrium electronic signal after cooling over hundreds of ps should become negligible in comparison to the ps region. However, we find that the photo-reflectivity of Cd$_2$Os$_2$O$_7$ contrasts to this seemingly trivial expectation.

In this articel, we present the non-equilibrium dynamics of Cd$_2$Os$_2$O$_7$. In the antiferromagnetic state, the photo-reflectivity initially shows the ps relaxation of the electrons. However, to our surprise, the photo-reflectivity increases again during the successive electron cooling process. In particular, near $T_N$, the photo-reflectivity after heating up the total system by 1 K in the nanosecond (ns) region is even larger than the initial value due to electron heating by more than 300 K. The huge resurgent behavior in the ns region suggests that an effective-temperature of another subsystem has an exceptionally strong influence on the transient electronic structure. We discuss our observation in terms of the local magnetic moment in the process of the Lifshitz transition. 

We measured the near-infrared photo-reflectivity as a function of the pump and probe delay time ($t_d$) on Cd$_2$Os$_2$O$_7$ single crystals grown by the chemical transport method \cite{Hiroi2015}. For the pump-probe measurements, we employed a commercial Ti:sapphire amplifier system, which generates pulses with a center energy of 1.55 eV and duration of 36 fs at a repetition rate of 250 kHz. For all measurements, the pump and probe fluences are set to 50 and 8.3 ${\mu}$J/cm$^2$, respectively. The spot sizes (full width half maximum) of the pump and probe beams are 84 and 44 ${\mu}$m, respectively. The pump and probe polarizations are perpendicular to each other to prevent the detection of scattered pump photons.

\begin{figure}[]
	\includegraphics[width = 0.45\textwidth]{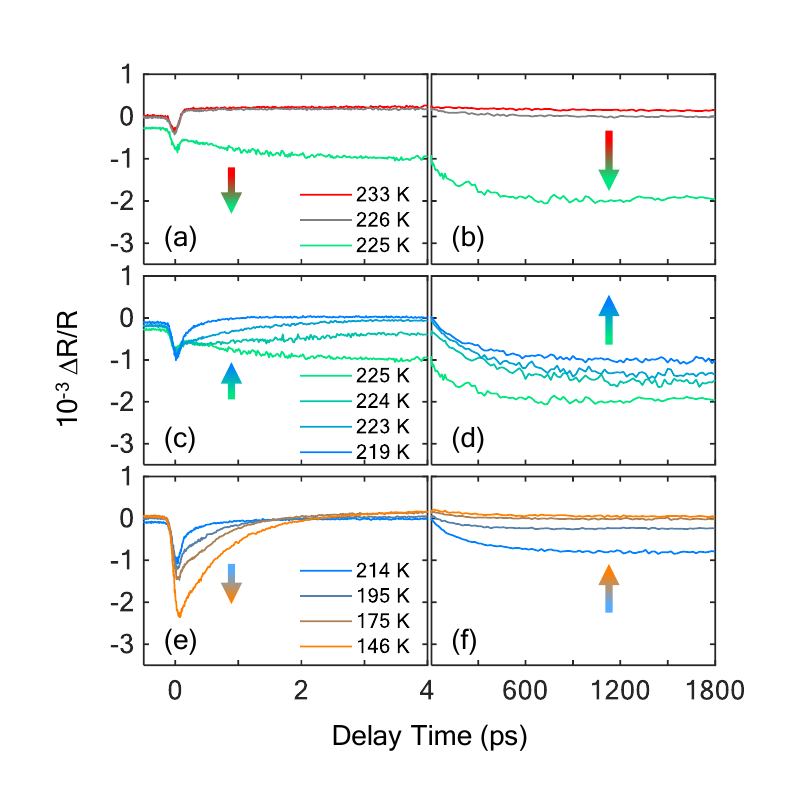}
	\caption{\label{fig:raw} Photo-reflectivity ${\Delta}R/R$ in two temporal windows, spanning -0.5 -- 4 ps (left panel), and 4 -- 1,800 ps (right panel) over temperatures from 146 K to 233 K.
	}
\end{figure}

Figure \ref{fig:raw} shows the temperature-dependent evolution of the photo-reflectivity ${\Delta}R/R$ in two time windows: -0.5 - 4 ps (left panel) and 4 - 1,800 ps (right panel). The ultrafast response above $T_N$ is typical of a metal, which exhibits a sub-ps relaxation followed by a small offset [Figs. \ref{fig:raw}(a) and \ref{fig:raw}(b)]. As already discussed in Fig. \ref{fig:sch}, the sub-ps relaxation corresponds to the hot electron cooling process via scattering with the SCPs \cite{Allen1987,Groeneveld1995,Perfetti2007,Giannetti2016,Averitt2002}. The finite offset lasting beyond the measurement time window is due to the heating of SCPs. Successive phonon-phonon scattering further cools down the electron and SCPs systems and also diffuses the excess heat out of the pump excited volume \cite{Perfetti2007}. 

Figures \ref{fig:raw}(c) and \ref{fig:raw}(e) show that additional, richly varied ps dynamics appears below $T_N$, and is much slower than the sub-ps dynamics that dominates above $T_N$. In the magnetic state, the electron-spin scattering may provide an additional cooling path for hot electrons. Once the effective-temperature of spin subsystem $T_S$ has reached to same temperature with $T_E$, however, successive cooling of hot electrons should be accompanied with the cooling of the spin subsystem. This results in the slower relaxation dynamics of Cd$_2$Os$_2$O$_7$ below $T_N$. Furthermore, the magnetic order could restrict the phase space of the electron-phonon scattering. This also explains the slower relaxation dynamics in the magnetic state. A more quantitative understanding of the ps dynamics in the magnetic state requires us to consider the spin subsystem as well as the electron and SCP subsystems in terms of the effective-temperatures \cite{Averitt2002,Patz2014}.

Unexpected is the result shown in Figs. \ref{fig:raw}(d) and \ref{fig:raw}(f) that the photo-reflectivity below $T_N$ resurges over hundreds of ps. The photo-reflectivity in the nanosecond (ns) region is so unusually large that it becomes comparable to or even larger than the initial photo-reflectivity. The large photo-reflectivity in the ns region shows up abruptly below $T_N$. Hence the magnetic order should contribute to this behavior. In fact, the spin-lattice relaxation is frequently observed in the ns region. When the spin-charge coupling is weak, however, the influence of $T_S$ on the electronic structure is marginal and the transient reflectivity does not show a significant change unless the heated magnetic state bears a resonance to the probe photon energy as in the case of Eu$_2$Fe$_2$As$_2$ \cite{Pogrebna2015}. The Os$^{5+}$ $d$-electron spins in Cd$_2$Os$_2$O$_7$ are known to be strongly coupled with electronic structure \cite{Shinaoka2012,Nguyen2017,Kim2016,Sohn2017}. Therefore, it is expected that the spin subsystem already plays a role in the ps region as discussed in the previous paragraph. In other words, $T_S$ as well as $T_E$ must monotonically decrease beyond 100 ps to reach the equilibrium state. Therefore, within the usual effective-temperature model, we cannot explain the huge resurgence in photo-reflectivity in the ns region in terms of $T_S$.

\begin{figure}[]
	\includegraphics[width = 0.45\textwidth]{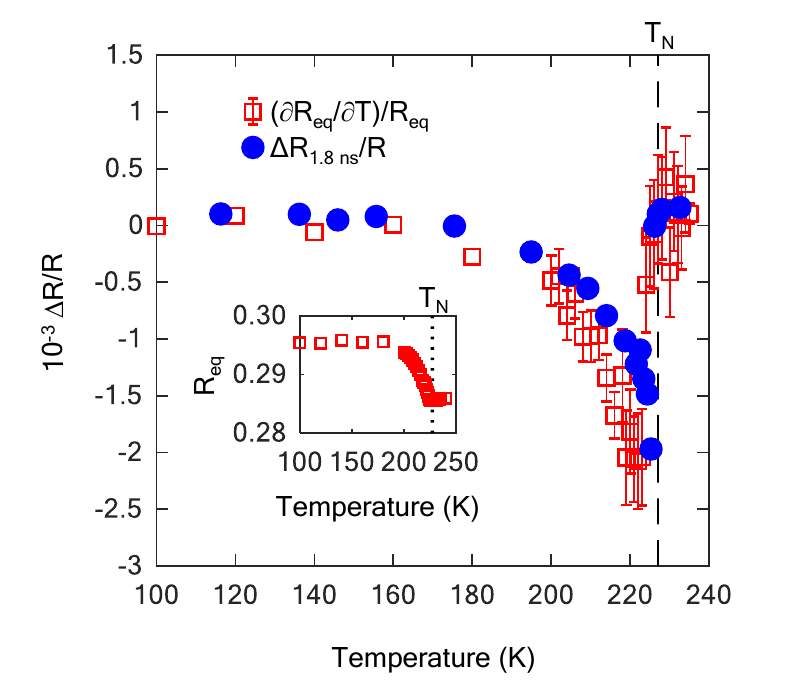}
	\caption{\label{fig:Req}(Color online) Temperature-dependent evolution of ${\Delta}R/R$ (circles) measured at $t_d$ = 1.8 ns and $({\partial}R_{\text{eq}}/{\partial}T)/R_{\text{eq}}$ (squares) obtained from the equilibrium reflectivity $R_{\text{eq}}$. Inset: The $R_{\text{eq}}$ at 1.55 eV measured by spectroscopic ellipsometry at 20 K intervals from 100 to 200 K, and at 1 K intervals from 200 to 235 K.
	}
\end{figure}

The simple thermodynamic simulation of the change in equilibrium reflectivity confirms that the resurgence in the reflectivity change can be attributed to the heating of the whole system. We assume that the injected pump energy is initially absorbed entirely by the electron subsystem. When the base temperature prior to pumping is close to $T_N$, we find that the pump excitation should heat up the electron subsystem by more than 300 K. After the heat has dissipated to all of the other subsystems, however, the temperature of the entire system is estimated to increase by only 1 K from the base temperature (see the Supplemental Material for further details \cite{Supple}). In Fig. \ref{fig:Req}, we compare the photo-reflectivity at $t_d$ = 1.8 ns (${\Delta}R_{1.8\text{ns}}/R$) with the thermo-modulation of the reflectivity $({\partial}R_{\text{eq}}/{\partial}T)/R_{\text{eq}}$. Note that the thermo-modulation corresponds to the change in reflectivity due to temperature increment by 1 K in the equilibrium state. The good agreement between ${\Delta}R_{1.8\text{ns}}/R$ and $({\partial}R_{\text{eq}}/{\partial}T)/R_{\text{eq}}$ suggests that the system indeed reaches a quasi-equilibrium state at $t_d$ = 1.8 ns, where the temperature of all subsystems is slightly elevated by 1 K, as shown in Fig. \ref{fig:sch}.

As ${\Delta}R_{1.8\text{ns}}/R$ can be explained in terms of the equilibrium reflectivity ($R_{\text{eq}}$), the large value of ${\Delta}R_{1.8\text{ns}}/R$ slightly below $T_N$ may seem trivial following from the temperature-dependent evolution in the equilibrium state. However, it is peculiar that small heating in the whole system induces a huge transient reflectivity comparable to or even larger than ${\Delta}R$ due to the electron heating. We note that the reflectivity is directly determined by the electronic structure of the material. As far as we know, all the reports based on the effective-temperature model have assumed that the ultrafast photo-reflectivity depends only on $T_E$ \cite{Allen1987,Groeneveld1995,Perfetti2007,Giannetti2016,Averitt2002}. Within the effective-temperature model, it has never been expected that the huge resurgence in photo-reflectivity may occur due to the tiny heating after completion of the ps relaxation to ${\Delta}R/R$ ${\sim}$ 0.

There are other examples of a large transient electronic response in the ns region. In superconductors, the photo-induced reflectivity change may increase over hundreds of ps under weak pump excitation \cite{Beck2013,Kabanov2005}. This slow response is due to the small superconducting pairing energy and can be well understood in terms of the quasiparticle dynamics. Some manganite systems also exhibit the relaxation dynamics over hundreds of ps with a large transient optical response, comparable to that observed in the ps region \cite{Lobad2001,Hirobe2005,Ren2008}. It has been suggested that manganite systems enter a photo-excited metastable state, which does not correspond to any equilibrium state at elevated temperatures \cite{Ren2008}. These are different from the case of Cd$_2$Os$_2$O$_7$. As far as we know, there is no other example, where the ns response comparable to the ps response in photo-reflectivity is induced by the tiny heating in the quasi-equilibrium state.

Cd$_2$Os$_2$O$_7$ is a unique system, in that its photo-reflectivity in the ns region is not explained only with $T_{E}$. In the non-equilibrium state, the heating of any subsystem may affect the photo-induced signal, such that 
\begin{equation} \label{eq:ETM}
	{\Delta}I = A_E {\Delta}T_E + A_{\text{SCP}} {\Delta}T_{\text{SCP}} + A_S {\Delta}T_S + A_L {\Delta}T_L,
\end{equation}
{\noindent}where ${\Delta}T_E$, ${\Delta}T_{\text{SCP}}$, ${\Delta}T_S$, and ${\Delta}T_L$ are the effective-temperature changes of electron, SCP, spin, and lattice subsystems. The coefficients $A_E$, $A_{\text{SCP}}$, $A_S$, and $A_L$ are evaluated based on what is measured as the photo-induced signal. Although these coefficients may depend on the equilibrium temperature, they can be regarded as constants under small temperature variation. Reflectivity is directly determined by the electronic structure and ${\Delta}T_E$ is much larger than the other effective-temperature changes right after pumping. The reflectivity change, therefore, has been considered to be dominated by the electron subsystem, such that ${\Delta}R/R \sim A_E {\Delta}T_E$. However, the resurgence in the photo-reflectivity of Cd$_2$Os$_2$O$_7$ cannot be explained only by ${\Delta}T_E$, because the second law of thermodynamics states that $T_E$ must continue to decrease after ps relaxation. To explain the resurgent behavior over hundreds of ps, another effective temperature other than $T_{E}$ should be considered. Note that ${\Delta}T_{\text{SCP}}$ and ${\Delta}T_{S}$ already play roles in the ps region as discussed with Figs. 2(c) and 2(e). Therefore, the remaining term is the total lattice system ${\Delta}T_L$ within Eq. (\ref{eq:ETM}). We note that the photo-reflectivity could be governed by other unknown subsystems not included in Eq. (\ref{eq:ETM}) instead of ${\Delta}T_L$. However, the dynamics shown in Figs. 2(d) and 2(f) is well described by a single exponential component and the effective temperatures of all subsystems are quasi-equilibrated in the ns region. Therefore, we may use ${\Delta}T_L$ to stand for the subsystem playing the important role. Considering ${\Delta}T_E^{\text{max}}$ ${\sim}$ 300 K, ${\Delta}T_{L}^{\text{max}}$ ${\sim}$ 1 K (see the Supplemental Material \cite{Supple}), and the value of ${\Delta}R_{1.8\text{ns}}/R$ comparable to the initial electronic response below $T_N$, we find that the additional effective temperature coined by ${\Delta}T_{L}$ has an extremely strong influence, such that ${\Delta}R/R \sim A_E {\Delta}T_E + A_L {\Delta}T_L$ with $A_L/A_E$ $\sim$ 300 near $T_{N}$.

Why should ${\Delta}T_L$ exert such a strong influence over the electronic structure of Cd$_2$Os$_2$O$_7$? What should be modified by ${\Delta}T_L$? Although the lattice shows a slight thermal expansion, our density functional theory calculations find that the effect of lattice expansion on the band structure is negligible (see the Supplemental Material for details of this calculation \cite{Supple}). Instead, it is expected that the local magnetic moment of osmium $m_{\text{Os}}$, especially the longitudinal fluctuations, which is not included in ${\Delta}T_{S}$, may play a certain role for the modification of electronic structure in the ns region \cite{Kim2016,Dong2017}. Note that the band structure of Cd$_2$Os$_2$O$_7$ strongly depends on the local magnetic moment of osmium, $m_{\text{Os}}$. The local spin density approximation calculations by Shinaoka \textit{et al.} revealed that Cd$_2$Os$_2$O$_7$ undergoes a Lifshitz-type MIT as the electron correlation strength $U$ varies \cite{Shinaoka2012}. This corresponds to that $m_{\text{Os}}$ decreases as temperature increases and induces the MIT as well as the AIAO magnetic transition in Cd$_2$Os$_2$O$_7$ \cite{Shinaoka2012,Chitra1999,Kim2016}. We conjecture that the longitudinal fluctuations of $m_{\text{Os}}$ may explain the observed slow dynamics coined by ${\Delta}T_L$. Unfortunately, it is still unveiled that how the heating of the total lattice modifies $m_{Os}$, which in turn strongly influence the electronic structure in the ns region. The temperature dependent subtle evolution of the effective correlation strength could be more important than a direct interaction between the lattice and  $m_{\text{Os}}$. Further studies are required to understand the coupling between the lattice temperature and $m_{\text{Os}}$ and its role on the Lifshitz transition in Cd$_2$Os$_2$O$_7$.

\begin{figure}[t]
	\includegraphics[width = 0.45\textwidth]{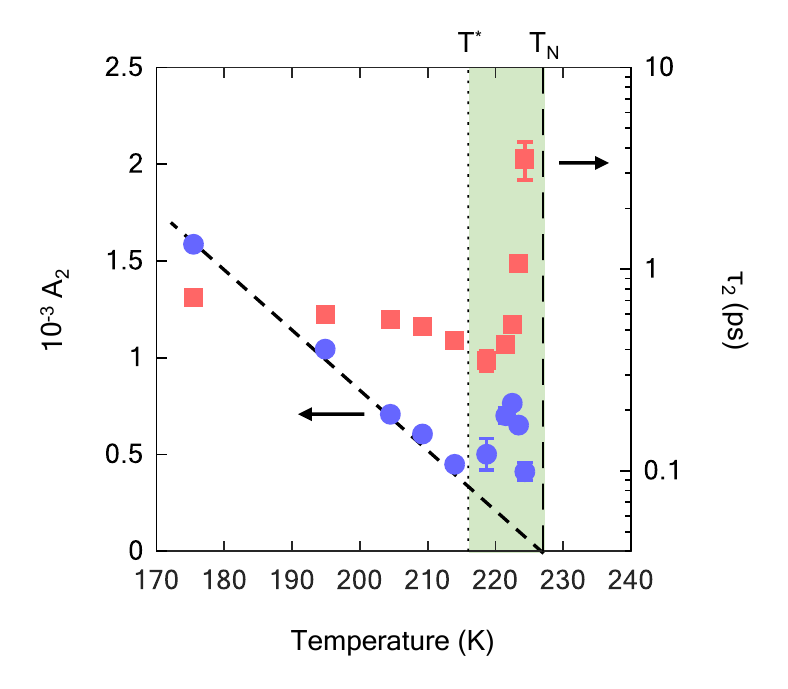}
	\caption{\label{fig:par}(Color online) Temperature dependence of the relaxation time ${\tau}_2$ (squares) and amplitude $A_2$ (circles) of the ps components below $T_N$, based on bi-exponential fitting to the photo-reflectivity ${\Delta}R/R$.
	}
\end{figure}

Finally, we briefly discuss the phase transition behaviors of the ps relaxation dynamics in Figs. \ref{fig:raw}(c) and \ref{fig:raw}(e). We obtained the temperature-dependence of the relaxation time (${\tau}_2$) and the amplitude ($A_2$) of the ps component that shows up below $T_{N}$ in detail by fitting our data to a bi-exponential decay model in the 10 ps region:
\begin{equation} \label{eq:fit}
	{\Delta}R/R(t_d)=A_1e^{-t_d/{\tau}_1}+A_2e^{-t_d/{\tau}_2}+C,
\end{equation}
{\noindent}where ${\tau}_1$ and $A_1$ are the relaxation time and amplitude of the sub-ps component, respectively, and $C$ is a constant offset. While the sub-ps component of ${\tau}_1$ ${\sim}$ 0.1 ps does not vary significantly, the ps component exhibits a strong temperature dependence, as shown in Fig. \ref{fig:par} (see the Supplemental Material for details of the fitting process \cite{Supple}). ${\tau}_2$ exhibits critical slowing down behavior as it approaches $T_N$ due to the AIAO phase transition \cite{Watanabe2009,Patz2014}. It %confirms 
suggests that the %transverse magnon 
excitation and recovery of the magnetic order with a given local magnetic moment $m_{\text{Os}}$ take place in this ps time window. The saturation behavior of the ps component under high fluence at low temperature well below $T_N$ also supports that the ps relaxation reflects a dynamics of an order parameter, while the ns one shows a linear response to the fluence (see the Supplemental Material for further details \cite{Supple}).

Figure 4 shows another anomaly in the ${\tau}_2$ and the $A_2$ at approximately $T^*$ ${\sim}$ 216 K. It has been claimed that this system exhibits a Lifshitz MIT at $T^*$ that is separate from the AIAO magnetic phase transition at $T_N$ \cite{Shinaoka2012,Sohn2015,Nguyen2017}. As mentioned above, the Lifshitz transition is driven by the correlation effect as a function of $m_{\text{Os}}$. In other words, the energy gap in the insulating state should open between the electron correlation-induced bands. In contrast to the slow electronic relaxation by recombination process in semiconductors and band gap insulators over more than 100 ps \cite{Sabbah2000,Sabbah2002}, the relaxation of correlated insulators has been reported to take place on the ps time scale, which is comparable to the relaxation of their metallic state \cite{Hsieh2012,Chu2017,Okamoto2010,Lee2018}. Therefore, the gradual evolution of the carrier relaxation dynamics across $T^*$ is consistent with the correlation-induced nature of the Lifshitz transition in Cd$_2$Os$_2$O$_7$.

In summary, we performed pump-probe reflectivity measurements on 5$d$ pyrochlore Cd$_2$Os$_2$O$_7$, a prototype material exhibiting the Lifshitz transition. The non-equilibrium dynamics reveals an exceptionally large resurgence in the photo-reflectivity, over hundreds of ps. %, as $T_L$ increases. 
It shows that a slight heating of the total system has a huge influence on the electronic structure much stronger than the heating of electrons.  
These findings are consistent with the Lifshitz-type transition controlled by the local magnetic moment of osmium $m_{\text{Os}}$ that depends sensitively on temperature. %given at $T_L$. 
The additional anomaly in the ps dynamics at $T^*$ ${\sim}$ 216 K supports that the Lifshitz-type MIT occurs below $T_N$. Our study of the prototypical response of the correlation-induced Lifshitz transition will provide a guide to understanding the ultrafast dynamics of correlated systems that exhibit emergent phenomena.

%\appendix
%\section{}

%%%%%%%%%%%%%%%%%%%%%%%%%%%%%%%%%%%%%%%%%%%%%%%%%%%%%%%%%%%%%%%%%%%%%%%%%%%%%%%%%%%%%%%%%%%%%%%%%%%

\begin{acknowledgments}
We thank S.Y. Kim, W. J. Kim for their experimental assistance and J. H. Sim, C. H. Sohn for valuable discussions. This work was supported by the Institute for Basic Science (IBS) in Korea (IBS-R009-D1) and by the Basic Science Research Program through the National Research Foundation of Korea (NRF) funded by the Ministry of Science, ICT and Future Planning (NRF-2017R1A4A1015323 and 2019R1F1A1062847).
\end{acknowledgments}

%%%%%%%%%%%%%%%%%%%%%%%%%%%%%%%%%%%%%%%%%%%%%%%%%%%%%%%%%%%%%%%%%%%%%%%%%%%%%%%%%%%%%%%%%%%%%%%%%%%

%

\makeatother
\newpage

\section*{Supplementary material}

1. **The maximum change of the effective temperature**

**(Homogeneous absorption approximation)**

By assuming that an injected energy of a pump pulse ($Q_{pump}$) is homogenously absorbed in the measuring region, we estimate the maximum change of the effective-temperatures of the electron ($\Delta T_E^{hom}$) and the lattice ($\Delta T_L^{hom}$).

To estimate $\Delta T_L^{hom}$, we employed the value of heat capacity $C_P$ of $Cd_2Os_2O_7$ single crystal from ref. [1]:

$$\Delta T_L^{hom} = \frac{\Delta Q_{pump}}{C_P}. \quad (S.\ 1)$$

To estimate $\Delta T_E^{hom}$, we calculate the Sommerfeld coefficient $\gamma$, the linear coefficient of electron heat capacity:

$$\gamma = \frac{\pi^2 m^* k_B^2}{\hbar^2} \left(3\pi^2 n\right)^{-2/3} (nV_m) \leq 3.0\ \text{mJ/(mol·K}^2). \quad (S.\ 2)$$

where $n$ is a carrier density, $V_m$ is a molar volume, and $m^*$ is an effective mass of an electron. To consider whole candidates of photo-excited carriers, we obtained $n$ by integrating optical conductivity at equilibrium temperature $T_{eq} = 225$ K from 0 eV to 1.55 eV photon energy [2]:

$$n \leq \frac{8\pi m^*}{h} \int_0^{1.55\ \text{eV}} \sigma(\omega) d\omega, \quad (S.\ 3)$$

$$\Delta Q_{pump} = \int_{T_{eq}}^{T_{eq}+\Delta T_E^{hom}} \gamma T\, dT = \frac{\gamma}{2}\left(2T_{eq}\Delta T_E^{hom} + \left(\Delta T_E^{hom}\right)^2\right), \quad (S.\ 4)$$

$$\Delta T_E^{hom} = T_{eq}\left(\sqrt{1+\frac{2\Delta Q_{pump}}{\gamma T_{eq}^2}} - 1\right). \quad (S.\ 5)$$

**(Exponential absorption & heat propagation approximation along depth)**

The assumption of the homogeneous absorption is not appropriate for estimation of the heating effect on optical properties from the maximum change of effective-temperatures ($\Delta T_E^{max}$ and $\Delta T_L^{max}$). Although the absorption at the sample surface is regarded to be homogeneous, the absorption along depth is not. Since the electromagnetic wave is exponentially attenuated with propagation, we should account for this by using the penetration depth twice (entering and exiting of the probe beam) [3]:

$$\Delta T^{max}(t_d) = \frac{2}{\lambda}\int_0^\infty \Delta T(z,t_d) \exp\left(-\frac{2z}{\lambda}\right) dz \quad (S.\ 6)$$

$$\Delta T(z,t_d) = T(z,t_d) - T_{eq} \quad (S.\ 7)$$

where $T(z,t_d)$ is the temperature distribution function at depth $z$ and at time $t_d$ after pumping, and $\lambda$ is the optical penetration depth. The value of $\lambda = 38$ nm at 1.55 eV is measured by ellipsometer.

To simulate $T(z,t_d)$ in the ns time domain, we employed a simple 1D heat equation with a boundary condition of the optical absorption:

$$\frac{\partial T(z,t_d)}{\partial t_d} - \alpha \frac{\partial^2 T(z,t_d)}{\partial z^2} = 0 \tag{S. 8}$$

$$T(z,0) = \Delta T^{\text{hom}} \exp\left(-\frac{z}{\lambda}\right) + T_{\text{eq}} \tag{S. 9}$$

where $\alpha$ is a thermal diffusivity. The values of $\alpha$ is employed from ref. [1]. Although this model is not realistic in the non-equilibrium state, it provides us a simple estimation of the heating effect in the quasi-equilibrium state. By solving 1D heat equation (S. 8) and (S. 9) numerically, we finally obtained $\Delta T_E^{\text{max}}$ and $\Delta T_L^{\text{max}}$ from (S. 6):

$$\Delta T_E^{\text{max}} \sim 333 \text{ K (at } t_d = 0 \text{ ns), and} \tag{S. 10}$$

$$\Delta T_L^{\text{max}} \sim 1 \text{ K (at } t_d = 1.8 \text{ ns).} \tag{S. 11}$$

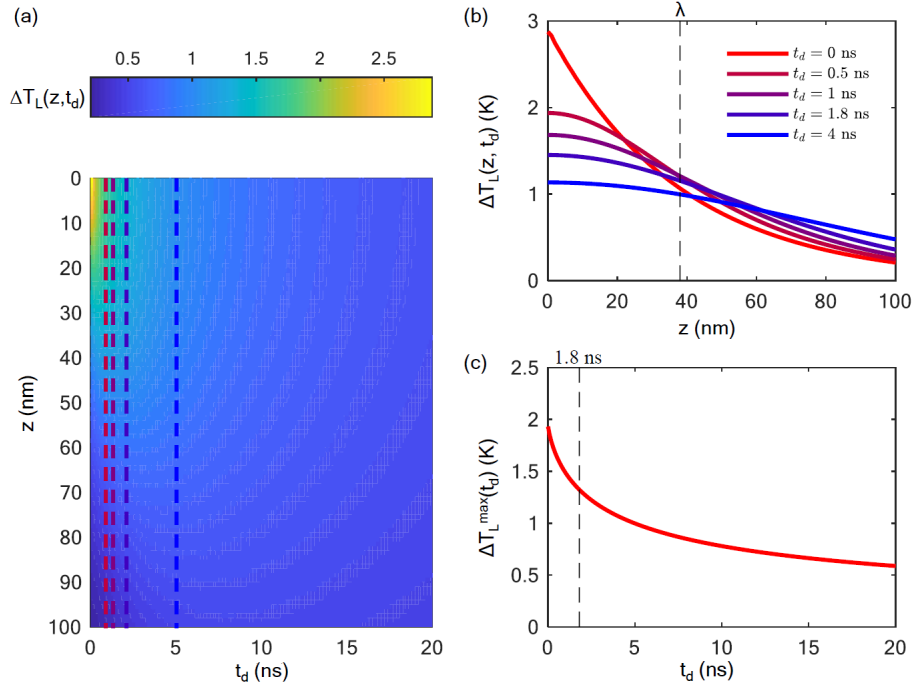

FIG. S1. (a) Temperature distribution function $T_L(z,t_d)$ at 100 nm and 20 ns window. (b) Lattice temperature change distribution $\Delta T_L(z,t_d)$ at various time delay $t_d$. (c) The lattice temperature change distribution $\Delta T_L(t_d)$ integrated over the depth.

## 2. Density functional theory band calculations

To identify influences of the lattice thermal expansion on the electronic structure, we performed density functional theory (DFT) band calculations including the electronic correlation $U$ of 0.7 eV. The structural symmetry and lattice constants at 180 K and 250 K are employed from ref. [4]. Despite the continuous expansion of the lattice across $T_N$, the calculated bands are nearly identical to each other. This result shows that the simple thermal expansion of the lattice cannot drive the significant band modification responsible for the resurgence of the photo-induced reflectivity change in $Cd_2Os_2O_7$.

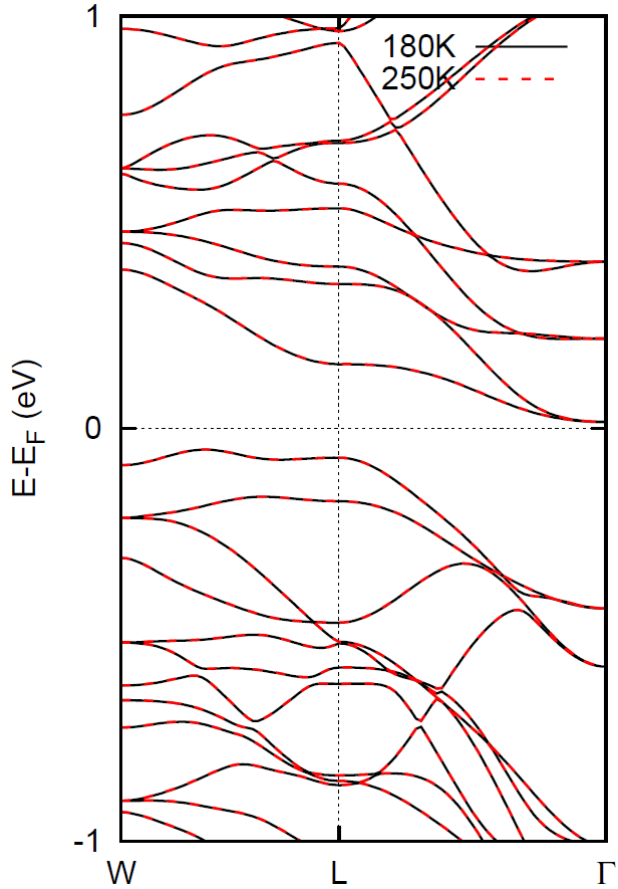

FIG. S2. The electronic band structure of $Cd_2Os_2O_7$ is calculated using DFT+$U$. The displayed band structure is calculated with $U = 0.7$ eV using reported structural information measured at 180 K (black solid lines) and 250 K (red dotted lines) [4]. They are nearly identical to each other.

## 3. Exponential decay fit analysis on the picosecond relaxation below $T_N$

Exponential fit analysis on the ps relaxation reveals remarkable features of the two sequential anomalies at all-in-all-out antiferromagnetic transition temperature $T_N = 227$ K and at $T^* \sim 216$ K. Figure S1 displays the experimental data (circle) and corresponding fitting lines (solid lines). The overall relaxation within 10 ps is well described by a bi-exponential decay model:

$$\Delta R/R(t_d) = A_1\, e^{-t_d/\tau_1} + A_2\, e^{-t_d/\tau_2} + C \quad \text{for } t_d \leq 10 \text{ ps} \tag{S. 8}$$

where $\tau_1$ ($\tau_2$) is the relaxation time, and $A_1$ ($A_2$) is the amplitude of the sub-ps (the ps) component respectively, and $C$ is the constant value. As displayed in Fig. 4(b), the ps components of $A_2$ and $\tau_2$ show significant temperature evolutions at phase transitions of $T_N$ and $T^*$. On the other hand, $\tau_1$ and $A_1$ of the sub-ps component do not show significant changes.

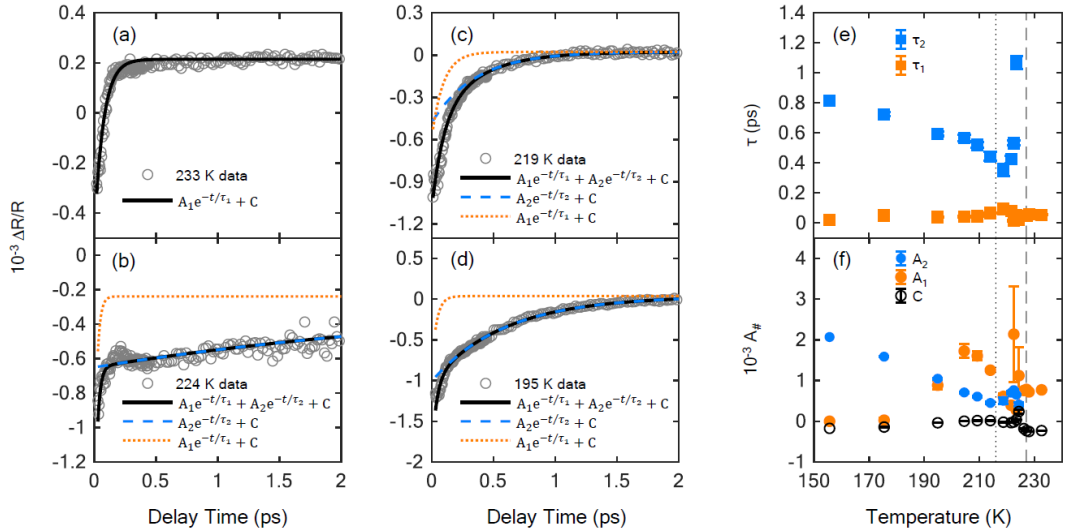

FIG. S3. (a-d) Fitting results on the data in the ps region which contains the sub-picosecond and the ps components. The experimental data (open circles) is overlaid with three fitting lines: bi-exponential fitting line (thick solid lines), the ps component (dashed lines), and the sub-ps component (dotted lines). (e-f) The obtained fitting parameters of $\tau_1$ (orange squares), $A_1$ (orange circles), and $C$ (black circles). For comparisons, $A_2$ (blue circles) and $\tau_2$ (blue squares) are displayed together.

## 4. Exponential decay fit analysis on the picosecond relaxation below $T_N$

We performed the transient reflectivity change measurements at 30 K far below $T_N$ with various pump fluences. The fluence-dependent evolution of photo-reflectivity obtained at 0.3 ps delay shows an exponential saturation behavior fitted by

$$\Delta R/R = A(1 - e^{-F/\Phi_{\text{sat}}}) \tag{S1}$$

where $\Phi_{\text{sat}} = 202 \pm 66$ $\mu$J/cm$^2$ is the saturation fluence. On the other hand, the fluence-dependent evolution at 0.04 ps and 10 ps delay times exhibit simple linear response. Our observations strongly support that the charge-spin coupling is so strong that the spin temperature rising and cooling process is relevant to the ps relaxation.

One may expect that the ps dynamics should exhibit the bi-recombination or Rothwarf-Taylor-like bottleneck dynamics explaining isotropic charge gap state. In such dynamics the relaxation time of the quasi-particle recombination across the gap strongly depends on the pump fluence. However, the ps relaxation time observed in Cd$_2$Os$_2$O$_7$ barely changes. Note that the insulating in Cd$_2$Os$_2$O$_7$ has correlation-induced indirect gap [1,5]. The understanding quasi-particle dynamics across such gap may require further extensive investigation.

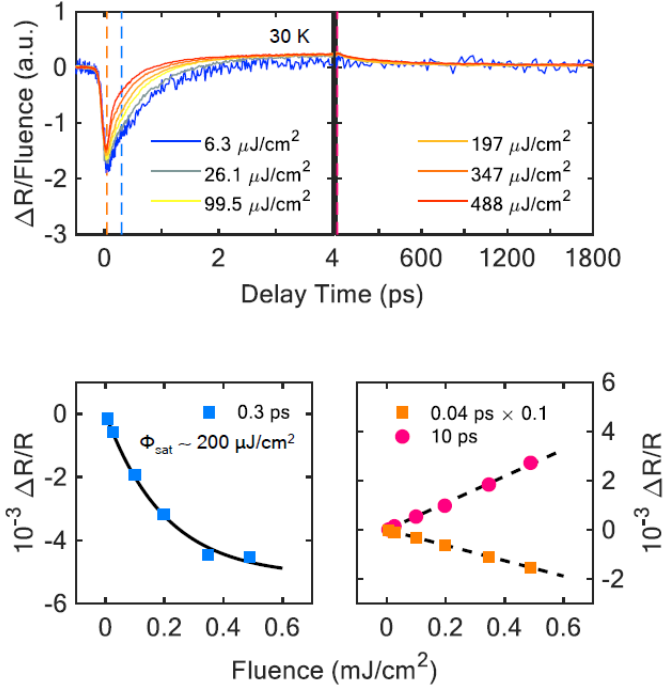

FIG. S4. Transient reflectivity change is measured at 30 K with various pump fluence. (a) Fluence-normalized transient reflectivity changes are displayed. (b) Fluence-dependent evolution of the transient reflectivity obtained at fixed delay times of (b) 0.3 ps, (c) 0.04 ps and 10 ps. The obtained values at 0.04 ps, 0.3 ps, and 10 ps corresponds respectively to the amplitudes of a sub-ps, the ps, and the hundreds ps relaxation components.




\begin{thebibliography}{}%
	
\bibitem{Lifshitz1960}%
	I. M. Lifshitz,
	Sov. Phys. JETP \textbf{11}, 1130 (1960).%
	
\bibitem{Shinaoka2012}%
	H. Shinaoka,
	T. Miyake,
	and S. Ishibashi, 
	Phys. Rev. Lett. \textbf{108}, 247204 (2012).%
	
\bibitem{Volovik2017}%
	G. E. Volovik,
	Fiz. Nizk. Temp. \textbf {43}, 57 (2017).%
	
\bibitem{Braganca2018}%
	H. Bragan\c{c}a,
	S. Sakai,
	M. C. O. Aguiar,
	and M. Civelli, 
	Phys. Rev. Lett. \textbf{120}, 067002 (2018).%
	
\bibitem{Wu2015}%
	Y. Wu, 
	N. H. Jo, 
	M. Ochi, 
	L. Huang, 
	D. Mou, 
	S. L.Bud'ko, 
	P. C. Canfield, 
	N. Trivedi, 
	R. Arita, 
	and A. Kaminski,
	Phys. Rev. Lett. \textbf {115}, 166602 (2015).%
		
\bibitem{Marsik2013}%
	P. Marsik \textit{et al.}, 
	Phys. Rev. B \textbf{88}, 180508(R) (2013).%
	
\bibitem{Ren2017}%
	M. Ren, 
	Y. Yan, 
	X. Niu, 
	R. Tao, 
	D. Hu, 
	R. Peng, 
	B. Xie, 
	J. Zhao, 
	T. Zhang, 
	and D.-L. Feng, 
	Sci. Adv. \textbf {3}, e1603238 (2017).%
	
\bibitem{Shi2017}%
	X. Shi \textit{et al.}, 
	Nat. Commun. \textbf {8},	14988 (2017).%
	
\bibitem{Chatterjee2017}%
	S. Chatterjee,
	J. P. Ruf,
	H. I. Wei,
	K. D. Finkelstein, 
	D. G. Schlom, 
	and K. M. Shen, 
	Nat. Commun. \textbf{8}, 852 (2017).%
	
\bibitem{Zhang2017}%
	Y. Zhang \textit{et al.}, 
	Nat. Commun. \textbf {8}, 15512 (2017).%

\bibitem{Padilla2002}%
	W. J. Padilla, 
	D. Mandrus, 
	and D. N. Basov,
	Phys. Rev. B \textbf{66}, 035120 (2002).%
	
\bibitem{Mandrus2001}%
	D. Mandrus, 
	J. R. Thompson, 
	R. Gaal,
	L. Forro,
	J. C. Bryan, 
	B. C. Chakoumakos, 
	L. M. Woods, 
	B. C. Sales, 
	R. S. Fishman, 
	and V. Keppens, 
	Phys. Rev. B \textbf{63}, 195104 (2001).%
	
\bibitem{Bogdanov2013}%
	N. A. Bogdanov, 
	R. Maurice,
	I. Rousochatzakis,
	J. van den Brink, 
	and L. Hozoi, 
	Phys. Rev. Lett. \textbf{110}, 127206 (2013).%
	
\bibitem{Calder2016}%
	S. Calder \textit{et al.}, 
	Nat. Commun. \textbf{7}, 11651 (2016).%

\bibitem{Nguyen2017}%
	T. M. H. Nguyen \textit{et al.},
	Nat. Commun. \textbf {8}, 251 (2017).%

\bibitem{Tardif2015}%
	S. Tardif,
	S. Takeshita,
	H. Ohsumi, 
	J.-i. Yamaura, 
	D. Okuyama, 
	Z. Hiroi, 
	M. Takata, 
	and T.-h. Arima,
	Phys. Rev. Lett. \textbf{114}, 147205 (2015).%
	
\bibitem{Yamaura2012}%
	J. Yamaura \textit{et al.}, 
	Phys. Rev. Lett. \textbf{108}, 247205 (2012).%

%\bibitem{Basov2011}%
%	D. N. Basov, 
%	R. D. Averitt,
%	D. van der Marel, 
%	M. Dressel, 
%	and K. Haule, 
%	Rev. Mod. Phys. \textbf{83}, 471 (2011).%
	
\bibitem{Beck2013}%
	M. Beck, 
	I. Rousseau,
	M. Klammer, 
	P. Leiderer, 
	M. Mittendorff, 
	S. Winnerl, 
	M. Helm, 
	G. N. Gol'tsman, 
	and J. Demsar, 
	Phys. Rev. Lett. \textbf{110}, 267003 (2013).%

\bibitem{Kabanov2005}%
	V. V. Kabanov, 
	J. Demsar, 
	and D. Mihailovic, 
	Phys. Rev. Lett. \textbf{95}, 147002 (2005).%

\bibitem{Allen1987}%
	P. B. Allen, 
	Phys. Rev. Lett. \textbf {59}, 1460 (1987).%

\bibitem{Groeneveld1995}%
	R. H. M. Groeneveld, 
	R. Sprik, 
	and A. Lagendijk, 
	Phys. Rev. B \textbf {51}, 11433 (1995).%

\bibitem{Perfetti2007}%
	L. Perfetti,
	P. A. Loukakos,
	M. Lisowski,
	U. Bovensiepen,
	H. Eisaki, and
	M. Wolf,
	Phys. Rev. Lett., \textbf{99}, 197001 (2007).%
	
\bibitem{Giannetti2016}%
	C. Giannetti, 
	M. Capone,
	D. Fausti, 
	M. Fabrizio, 
	F. Parmigiani, 
	and	D. Mihailovic, 
	Adv. Phys. \textbf{65}, 58 (2016).%
	
\bibitem{Averitt2002}%
	R. D. Averitt 
	and A. J. Taylor, 
	J. Phys. Condens. Matter \textbf {14}, R1357 (2002).%
	
\bibitem{Lisowski2005}%
	M. Lisowski, 
	P. A. Loukakos, 
	A. Melnikov, 
	I. Radu, 
	L. Ungureanu, 
	M. Wolf, and
	U. Bovensiepen,
	Phys. Rev. Lett. \textbf{95}, 137402 (2005).
	
\bibitem{Watanabe2009}%
	S. Watanabe, 
	R. Kondo, 
	S. Kagoshima, 
	and	R. Shimano, 
	Phys. Rev. B \textbf{80}, 220408(R) (2009).%

\bibitem{Patz2014}%
	A. Patz,
	T. Li, 
	S. Ran,
	R. M. Fernandes, 
	J. Schmalian, 
	S. L. Bud'ko, 
	P. C. Canfield, 
	I. E. Perakis, 
	and J. Wang, 
	Nat. Commun. \textbf{5}, 3229 (2014).%
	
\bibitem {Gerber2017}%
	S. Gerber \textit{et al.}, 
	Science \textbf{357}, 71 (2017).%
	
\bibitem{Smallwood2012}%
	C. L. Smallwood, 
	J. P. Hinton, 
	C. Jozwiak,
	W. Zhang, 
	J. D. Koralek, 
	H. Eisaki, 
	D.-H. Lee, 
	J. Orenstein, 
	and	A. Lanzara, 
	Science \textbf{336}, 1137 (2012).%
	
\bibitem{Kim2012}%
	K. W. Kim, 
	A. Pashkin,
	H. Sch\"{a}fer, 
	M. Beyer, 
	M. Porer, 
	T. Wolf, 
	C. Bernhard, 
	J. Demsar,
	R. Huber, 
	and A. Leitenstorfer, 
	Nat. Mater. \textbf{11}, 497 (2012).%
	
\bibitem {Gerber2015}%
	S. Gerber \textit{et al.}, 
	Nat. Commun. \textbf{6}, 7377 (2015).%

\bibitem{Hiroi2015}%
	Z. Hiroi, 
	J. Yamaura,
	T. Hirose, 
	I. Nagashima, 
	and Y. Okamoto, 
	APL Mater. \textbf {3}, 041501 (2015).%
	
%\bibitem {Fann1992}%
%	W. S. Fann, 
%	R. Storz,
%	H. W. K. Tom, 
%	and	J. Bokor, 
%	Phys. Rev. Lett. \textbf{68}, 2834 (1992).%

%\bibitem{Guo2001}%
%	C. Guo, 
%	G. Rodriguez, 
%	and A. J. Taylor, 
%	Phys. Rev. Lett. \textbf {86}, 1638 (2001).%
	
%\bibitem{Cheng2017}%
%	L. Cheng,
%	Q.-B. Yan, and
%	M. Hu,
%	Phys. Chem. Chem. Phys., \textbf{19}, 21714 (2017).%

\bibitem{Pogrebna2015}%
	A. Pogrebna,
	T. Mertelj,
	N. Vuji\v{c}i\'{c},
	G. Cao,
	Z. A. Xu,
	and D. Mihailovic,
	Sci. Rep. \textbf {5}, 7754 (2015).%
	
\bibitem{Kim2016}%
	B. Kim,
	P. Liu,
	Z. Erg\"{o}nenc,
	A. Toschi,
	S. Khmelevskyi,
	and C. Franchini,
	Phys. Rev. B \textbf{94}, 241113(R) (2016).%

\bibitem{Sohn2017}%
	C. H. Sohn \textit{et al.},
	Phys. Rev. Lett. \textbf{118}, 117201 (2017).%
	
\bibitem{Supple}%
	Please see the Supplemental Material for additional experimental and theoretical details.%

\bibitem{Lobad2001}%
	A. I. Lobad, 
	R. D. Averitt,
	and A. J. Taylor, 
	Phys. Rev. B \textbf {63}, 060410(R) (2001).%

\bibitem{Hirobe2005}%
	Y. Hirobe, 
	Y. Kubo, 
	K. Kouyama, 
	H. Kunugita, 
	K. Ema, 
	and H. Kuwahara, 
	Solid State Commun. \textbf {133}, 449 (2005).%

\bibitem{Ren2008}%
	Y. H. Ren, 
	M. Ebrahim,
	H. B. Zhao, 
	G. L\"{u}pke, 
	Z. A. Xu, 
	V. Adyam, 
	and Q. Li, 
	Phys. Rev. B \textbf{78}, 014408 (2008).%

\bibitem{Dong2017}%
	Z. Dong,
	W. Li,
	D. Chen,
	S. Sch\"{o}necker,
	M. Long,
	and L. Vitos,
	Phys. Rev. B \textbf{95} 054426 (2017).%

\bibitem{Chitra1999}%
	R. Chitra 
	and G. Kotliar, 
	Phys. Rev. Lett. \textbf{83}, 2386 (1999).%

\bibitem{Sohn2015}%
	C. H. Sohn \textit{et al.}, 
	Phys. Rev. Lett. \textbf{115}, 266402 (2015).%

\bibitem{Sabbah2000}%
	A. J. Sabbah 
	and D. M. Riffe, 
	J. Appl. Phys. \textbf{88}, 6954 (2000).%

\bibitem {Sabbah2002}%
	A. J. Sabbah 
	and D. M. Riffe, 
	Phys. Rev. B \textbf{66}, 165217 (2002).%

\bibitem{Hsieh2012}%
	D. Hsieh, 
	F. Mahmood,
	D. H. Torchinsky,
	G. Cao, 
	and N. Gedik, 
	Phys. Rev. B \textbf{86}, 035128 (2012).%

\bibitem{Chu2017}%
	H. Chu, 
	L. Zhao, 
	A. de la Torre, 
	T. Hogan, 
	S.D. Wilson, 
	and D. Hsieh, 
	Nat. Mater. \textbf{16}, 200 (2017).%

\bibitem{Okamoto2010}%
	H. Okamoto, 
	T. Miyagoe,
	K. Kobayashi, 
	H. Uemura, 
	H. Nishioka, 
	H. Matsuzaki, 
	A. Sawa, 
	and Y. Tokura,
	Phys. Rev. B \textbf{82}, 060513(R) (2010).%

\bibitem{Lee2018}%
	M.-C. Lee \textit{et al.}, 
	Phys. Rev. B \textbf{98}, 161115(R) (2018).%

\end{thebibliography}
\end{document}